\documentclass[preprint]{aastex}
\shorttitle{Globular Clusters in NGC 4365}
\shortauthors{Brodie et al.}
\def\etal{{\it et al.}}

\begin{document}

\title{Old Globular Clusters Masquerading as Young in NGC 4365?}

\author{Jean P. Brodie, Jay Strader, Glenda Denicol\'o, Michael A. Beasley and A. J. Cenarro}
\affil{UCO/Lick Observatory, University of California, Santa Cruz, 95064}
\email{brodie@ucolick.org, strader@ucolick.org, denicolo@tlaloc.inaoep.mx, mbeasley@ucolick.org, and cen@astrax.fis.ucm.es}

\and

\author{S\o ren S. Larsen and Harald Kuntschner}
\affil{European Southern Observatory, Karl-Schwarzschild-Str. 2, D-85748, Garching bei M\"unchen, Germany}
\email{slarsen and hkuntsch @eso.org}

\and

\author{Duncan A. Forbes}
\affil{Centre for Astrophysics \& Supercomputing, Swinburne University, Hawthorn VIC 3122, Australia}
\email{dforbes@astro.swin.edu.au}

\begin{abstract}

High signal-to-noise, low-resolution spectra have been obtained for 22 globular clusters (GCs) in NGC~4365. Some of these were selected as
probable representatives of an intermediate-age (2--5 Gyr), extremely metal-rich GC subpopulation. The presence of such a subpopulation had
been inferred from the unusual optical and near-IR color distributions of GCs in this otherwise typical Virgo elliptical galaxy. However,
ages derived from Lick indices are consistent with uniformly old mean ages for all GCs in our sample. The metallicities of the clusters
show some evidence for a \emph{trimodal} distribution. The most metal-poor and metal-rich peaks are consistent with the values expected for
an elliptical galaxy of this luminosity, but there appears to be an additional, intermediate-metallicity peak lying between them. New
\emph{Hubble Space Telescope} photometry is consistent with this result. A plausible scenario is that in earlier data these three peaks
merged into a single broad distribution. Our results suggest that it is difficult to identify intermediate-age GC subpopulations solely
with photometry, even when both optical and near-infrared colors are used.

\end{abstract}

\keywords{globular clusters: general -- galaxies: individual (NGC 4365) -- galaxies: stellar content -- galaxies: evolution}

\section{Introduction}

Several studies have demonstrated that the color distributions of the GC systems of luminous early-type galaxies are usually found to be
bimodal---the color distribution is better represented by a superposition of two Gaussian functions than by a single Gaussian, typically
with a high level of statistical significance (Ashman, Bird, \& Zepf 1994; Gebhardt \& Kissler-Patig 1999; Kundu \& Whitmore 2001; Larsen
\etal~2001). The vast majority of these studies have been carried out in the $V$ and $I$ passbands and the peaks of the Gaussians occur
consistently at or around $V-I\sim0.95$ and $V-I\sim1.18$. This almost ubiquitous bimodality indicates the common presence of
subpopulations, and suggests that multiple mechanisms and/or epochs of formation have been at work in producing the GC systems and, by
implication, the parent galaxies themselves. Since there is an age-metallicity degeneracy in optical broadband colors, it is not possible
to determine from optical photometry alone whether the color difference is due to an age or a metallicity difference between the
subpopulations or some combination of the two. If both subpopulations are old, like those in the Milky Way ($\sim 10-13$ Gyr), the color
difference would correspond to a metallicity difference of $\sim 1$ dex in [Fe/H]. In general, a younger subpopulation will appear bluer
for a given metallicity.

The giant elliptical galaxy NGC 4365, located in the outskirts of the Virgo Cluster, has previously been noted as a puzzling possible
exception to the bimodality rule. Its HST/WFPC2 $V-I$ color distribution is unimodal and broad (Forbes 1996; Gebhardt \& Kissler-Patig
1999; Larsen \etal~2001). Hundreds of GCs were detected in these studies, indicating that the unimodality cannot be attributed to small
number statistics. Puzia \etal~(2002) suggested, based on a combination of optical and near-infrared photometry, that a significant
fraction of the GCs in NGC 4365 are of intermediate-age (2--5 Gyr) and are extremely metal-rich (between solar and 3 times solar
metallicity). Such intermediate-age clusters might fill in the gap between the blue and red peaks in a ``normal'' color distribution,
causing it to appear unimodal and unusually wide for a single subpopulation. Using Keck spectroscopy, Larsen \etal~(2003, hereafter Paper
I) estimated ages and metallicities for a sample of 14 confirmed GCs and found evidence that some, although not all, of the candidate
intermediate-age clusters were intermediate-age and metal-rich ($-0.4 \la\,$[Fe/H]$\,\la 0$). Interestingly, the derived metallicities were
significantly lower than those determined by Puzia \etal~on the basis of the extremely red $V-K$ colors of the clusters.

In the context of GCs as probes of the formation histories of galaxies, these results presented something of a contradiction. While early
work suggested stellar population differences between the main body of the galaxy and its kinematically decoupled core (Surma \& Bender
1995), a more recent study found a luminosity-weighted age of $\sim 14$ Gyr and constant [Mg/Fe] for both components (Davies \etal~2001).
NGC 4365 shows no evidence for a recent merger. If a significant subpopulation of intermediate-age GCs is indeed present in this galaxy,
where are the intermediate-age galaxy stars that should accompany it? In Paper I we pointed out that at most 2-3\% of the mass of the
galaxy could be in a stellar population with an age of 2--5 Gyr without strongly affecting its old integrated spectrum.

Here we revisit the NGC~4365 GC system with a new Keck spectroscopic study which yielded 22 confirmed GCs, three of which are in common
with Paper I. The new sample is of higher signal-to-noise (S/N) than the previous work. Diagnostic absorption line indices have been
measured on these new spectra and compared with the most recent simple stellar population (SSP) models to derive ages and metallicities for
the individual GCs. New HST/ACS data for NGC 4365, obtained in the $g$ and $z$ bands, provides a better baseline for metallicity
determinations than the WFPC2 $V-I$ data used in previous studies. The larger sample, superior data quality, and more sophisticated stellar
evolutionary synthesis models allow a more reliable estimate of the distribution of ages and metallicities in the NGC~4365 GC system.

In \S 2 we discuss the observations and data reduction techniques. In \S 3 we derive radial velocities, ages, metallicities, and
[$\alpha$/Fe] ratios for the GCs in our sample. In \S 4 we present the analysis of HST/ACS $g$ and $z$ photometry for NGC 4365. \S 5 has
discussion and conclusions.

\section{Observations and Data Reduction}

The spectra were obtained in multi-slit mode on 2003 March 23 and 24 using the LRIS spectrograph (Oke \etal~1995) on the Keck I telescope.
Candidate GCs were selected from $V$ and $I$ preimaging from FORS1 on VLT/UT1. Exposure times were $3\times140$ s in $V$ and $3\times190$ s
in $I$. The FORS photometry was roughly calibrated using HST/WFPC2 data for NGC 4365 (Larsen \etal~2001; Kundu \& Whitmore 2001). GC
candidates were selected on the basis of luminosity and spatial distribution, covering a range of galactocentric radii out to $1.5~r_e$
(Figure 1). Both blue and red candidates were selected, and we specifically chose several of the possible intermediate-age GCs identified
in Puzia \etal~(2002). The sample included three objects from our previous study: our 6, 8, and 13 are numbers 2, 5, and 11 in Paper I.
During the observations, the slitmask was aligned at a position angle of 25$^\circ$. 17 exposures, each with an integration time of 30
minutes, yielded a total exposure time of 510 min (8.5 hours) for each GC.

The blue side of LRIS, with a 600 l/mm grism, provided a wavelength coverage of $\sim 3800-5600$ \AA. Flux standards G191-B2B, Hiltner 600,
and PG0934+554, and Lick/IDS standard stars from Worthey \etal~(1994), were observed. The Lick standard stars were also used as radial
velocity standards with kinematic information extracted from SIMBAD\footnote{SIMBAD: operated at CDS, Strasbourg, France.}. The slitlets in
the mask employed on the first night were 0.8$\arcsec$ wide, matching the excellent seeing. The 1.0$\arcsec$ slitmask was chosen for the
second night during which the seeing varied between 1.0-1.2$\arcsec$. Relevant properties for the cluster candidates are listed in Table 1.
The $V$ photometry is from our FORS imaging.

All the data reduction was carried out under IRAF\footnote{IRAF is distributed by the National Optical Astronomical Observatories, which
are operated by the Association of Universities for Research in Astronomy, Inc. under contract with the National Science Foundation.}. The
images were bias-subtracted and flatfield corrected. Cosmic rays were carefully removed from each frame. After correction for optical
distortions, each individual spectrum was extracted.  Following wavelength calibration based on arc lamps mounted within the spectrograph,
flux calibration was applied and the individual spectra of each globular cluster candidate were combined to optimize the S/N. After
measuring the radial velocities (see below), small corrections to the wavelength calibration were made by cross-correlating the coadded
spectra with appropriate SEDs from the new MILES stellar population models (Vazdekis \etal~2005).

\section{Kinematics, Ages, and Metallicities}

\subsection{Radial Velocities}

Radial velocities were determined by cross-correlating the target spectra with velocity standards. The radial velocities listed in Table 1
are an average of the cross-correlation results for each reference star; the errors are estimates of the standard error of the mean based
on 6 measurements. Of 24 candidate GCs, one object was found to be a foreground star and a second object appears to be a $z \sim 0.07$
emission-line galaxy, leaving 22 confirmed GCs. These have a mean radial velocity of 1143 $\pm$ 67 km/s and a dispersion of 315 km/s,
consistent with the RC3 value of $1227 \pm 13$ km/s (de Vaucouleurs \etal~1993). The GC subpopulations (see below) do not show significant
differences in their mean velocities or dispersions.

GC 4 shows H$\beta$ fill-in and [\ion{O}{3}] emission at 4959 and 5007 \AA, suggesting that it contains a planetary nebula. We defer
discussion of this cluster to a future paper and do not include it in our subsequent analysis.

Smoothed spectra for three of our sample GCs are shown in Figure 2.

\subsection{Lick Indices and Ages}

Lick indices were measured using the definitions in Trager \etal~(1998) and Worthey \& Ottaviani (1997) after using a wavelength-dependent
Gaussian kernel to smooth the final combined spectra to the Lick resolution. Errors were calculated statistically as the standard error of
the mean of indices of the individual spectra of each GC candidate, and checked by bootstrapping. The index measurements and associated
errors are given in Tables 2 and 3.  Measurements of Lick indices on six standard stars showed no significant deviations from the Lick
system (though errors in offset determinations were large compared to the deviations themselves), and offsets generally smaller than the
GC index errors. Thus, we did not apply them to our indices, but we have included them in Table 4 for completeness.

The Lick index values were compared with SSP model predictions to estimate ages and metallicities for the GCs. The models of Thomas
\etal~(2004, TMK04) were selected for this comparison. The TMK04 models are based on the ``$\alpha$-enhanced'' models of Thomas
\etal~(2003, TMB03) but they incorporate the $\alpha$-dependence of the higher order Balmer lines (H$\gamma$ and H$\delta$) in addition to
the H$\beta$ dependence included in TMB03. These models are the only ones currently available that explicitly account for the
$\alpha$-enhancement known to dominate the abundance patterns in MW GCs and thus expected to be present in extragalactic GCs.

Figures 3--5 show TMK04 SSP model grids for [$\alpha$/Fe] = +0.3 along with Balmer line indices (H$\beta$, H$\gamma_F$ and H$\delta_F$)
vs.~[MgFe]$\arcmin$ for the 21 GCs in our new sample. [MgFe]$\arcmin$ (defined in TMB03 as $\sqrt{\textrm{Mg}{\it
b}\,(0.72\,\textrm{Fe5270+0.28\,Fe5335})}$ varies only weakly with [$\alpha$/Fe]. Weighted means of the three subpopulations (see below)
are overplotted. The distribution of points in Figures 3--5 is consistent with a uniformly old ($\sim 10-13$ Gyr) GC population with a
spread of metallicity from [$Z$/H] $\sim -1.35$ to just above solar, though none appear to be far above solar. All three of the Balmer line
indices present a largely consistent picture of old ages for all clusters, though individual cluster ages cannot be determined accurately.
Ages derived from H$\gamma_F$ for the metal-rich GCs are slightly younger. Using the wider H$\gamma_A$ and H$\delta_A$ index definitions
instead of H$\gamma_F$ and H$\delta_F$ gives consistent results.

A feature obvious in Figure 5 (and noticeable, although less clearly, in Figures 3 and 4) is the grouping of objects into three distinct
areas, suggesting the presence of \emph{three} subpopulations of clusters. The weighted subpopulation means are consistent with this
interpretation. All galaxies which have been spectroscopically studied to date have been found to contain at most two old subpopulations of
GCs. Additional evidence for these three subpopulations, which we term metal-poor (blue), intermediate-metallicity (orange), and metal-rich
(red), is explored more fully in \S 4 using new HST/ACS photometry.

For a more quantitative analysis of GC ages we used the $\chi^2$ minimization technique of Proctor, Forbes, \& Beasley (2004), which
utilizes the entire suite of Lick indices to derive ages and metallicities for each GC. For most clusters all of the indices through Fe5406
were used, excepting CN$_1$ and CN$_2$. The resulting mean subpopulation ages are---blue: $11 \pm 1$ Gyr, orange: $11 \pm 2$ Gyr, red: $10
\pm 1$ Gyr. Systematic errors could affect these ages and thus they should not be taken as absolute. They do, however, illustrate that the
three subpopulations are coeval within the errors.

In Figures 3--5, the location of the galaxy itself in the model grids is indicated by a star symbol. Index values for the galaxy are taken
from Howell \& Guhathakurta (2004) for all indices except H$\delta_F$, which is from Denicol\'o \etal~(2005). These points represent an
$r_e$/8 aperture at the galaxy center. The galaxy shows no detectable [\ion{O}{3}], so no emission correction was required (however,
H$\alpha$ is not included in our spectra). Since Howell \& Guhathakurta (2004) find [$\alpha$/Fe] = +0.26 $\pm 0.04$ for NGC 4365, our use
of [$\alpha$/Fe] = +0.3 model grids is reasonable. The age of the galaxy is $\sim$ 7 and 4 Gyr derived from H$\beta$ and H$\gamma_F$, but
$\sim 13$ Gyr from H$\delta_F$. The 7 Gyr age we derive from H$\beta$ is $\sim 5$ Gyr younger than that reported by Davies \etal~(2001),
though our H$\beta$ index measurements are quite similar. The difference in the age estimate is entirely due to model differences; Davies
\etal~used the models of Vazdekis (1999) while we use the models of TMK04. This discrepancy nicely illustrates the fact that SSP models do
not offer reliable absolute age estimates. The relative rankings of galaxy and GCs in age and metallicity should be more robust, modulo
such factors such as horizontal branch morphology, the luminosity function of the red and asymptotic giant branches, and the difficulty of
model calibration at super-solar metallicities. These effects remain as uncertainties in SSP models (e.g., Schiavon \etal~2004; Maraston
\etal~2001).

While the center of NGC 4365 does appear to be slightly younger than its GC system, it is important to remember that the diagnostic indices
are luminosity-weighted. Thus, the interpretation of the H$\beta$ 7 Gyr age estimate could either be taken at face value, or could reflect
a very small amount (in terms of the mass involved) of recent star formation, superimposed on an underlying ancient stellar population
which entirely dominates the mass fraction. To quantify the latter possibility, we first note that the H$\beta$ value (at fixed
metallicity) of a 13 Gyr TMK04 stellar population is $\sim 1.37$, implying a difference of $\sim 0.18$ in index strength between 7 and 13
Gyr. Using Worthey (1994) models at solar metallicity, only $4\%$ of 2 Gyr light (or $1\%$ of 1 Gyr light) mixed with a 13 Gyr underlying
population would be needed to produce the observed effect.

\subsection{Comparison to Previous Work}

In Paper I the measured spectral indices showed substantial scatter along the locus of the age-metallicity degeneracy. The indices of some
GCs (particularly H$\beta$) fell below the SSP grid, formally suggesting very old ages. This is a well known problem with GC indices (e.g.,
Proctor \etal~2004). At the other extreme, some GCs occupied the intermediate-age, high-metallicity region of the model grid, exactly as
expected from the photometric work of Puzia \etal~(2002). However, it appears that this was largely due to relatively low S/N spectra (more
than $50\%$ lower than the present work), with errors causing GCs to spread out along the age-metallicity locus. Some lower S/N spectra
appear to have been affected by sky subtraction difficulties because of superposition on the bright, strongly spatially varying galaxy
background. Independent, careful re-reductions of the original dataset gave index values consistent within the errors of those published in
Paper I. However, some of these index values differ significantly (by 1 \AA\ or more) from the indices measured on the three overlapping
objects in this paper, indicating systematic uncertainties in addition to the formal (photon-statistical) errors. These results reinforce
the need for high S/N data, at as high a resolution as is possible, for studies of extragalactic GCs.

\subsection{Metallicities}

We derived metallicities for our sample clusters using PCA-based metallicities (Strader \& Brodie 2004) for a subset of 11 Lick indices as
well as the Proctor, Forbes, \& Beasley~(2004) $\chi^2$ minimization technique. These values can be found in Table 5. The typical formal
error on these PCA metallicities is 0.1 dex, but the systematic error may be 0.1--0.2 dex, or even larger if the cluster is not in the
metallicity range $-1.7 <$ [m/H] $< 0$ (see Strader \& Brodie 2004 for more details). Typical $\chi^2$ [m/H] errors are 0.2--0.3 dex. The
mean [m/H] values and sigmas of each of the three subpopulations for the PCA and $\chi^2$ methods are---blue: $-1.02\pm0.17$,
$-0.98\pm0.13$; orange: $-0.45\pm0.12$, $-0.41\pm0.14$; red: $-0.09\pm0.18$, $-0.13\pm0.11$. These mean metallicities are consistent, both
with each other and with the values qualitatively apparent in Figures 3--5.

\subsection{Abundance ratios}

In Figure 6 we compare our measurements of Mg$b$ against $<$Fe$>$ with 3 and 13 Gyr model isochrones from TMK04. The
intermediate-metallicity and metal-rich GCs in our sample have [$\alpha$/Fe] around +0.3 or higher. The metal-poor GCs have slightly lower
[$\alpha$/Fe] at $\sim +0.2$. Anomalously low abundance ratios (solar or lower) have been seen in metal-poor GCs in the elliptical NGC 3610
(Strader, Brodie, \& Forbes 2004) and in several spirals in the Sculptor Group (Olsen \etal~2004); these studies also used the TMK04
models. This may suggest a systematic problem with these models at low metallicities.

Figure 7 shows that the Lick CN$_2$ values of the NGC 4365 GCs are comparable to those of Galactic GCs of similar metallicity (which, in
turn, are enhanced with respect to local subdwarfs). However, the CN$_2$ index measures the $\Delta v = -1$ 4215 \AA\ band, which is rather
weak compared to the $\Delta v = 0$ 3883 \AA\ band. Indeed, M31 GCs are clearly enhanced in the UV band (attributed primarily to nitrogen,
Brodie \& Huchra 1991; Worthey 1998) but are quite similar to Galactic GCs in their CN$_2$ indices (Beasley \etal~2004), indicating the low
sensitivity of the Lick index to CN variations. At present, the typical CN enhancement for GCs is unclear.

\section{Photometry}

Figure 8 is a $V-I$ vs.~$V-K$ plot for GC candidates in NGC 4365 along with SSP model grids from Maraston (2005). GCs with Keck
spectroscopy (both from this work and from Paper I) are indicated by filled circles. The $VIK$ data are from Puzia \etal~(2002), who
suggested that the distribution of GCs in this color-color diagram reflected the presence of an intermediate-age (ranging from 2--5 Gyr)
subpopulation with solar to super-solar metallicities. Clearly the spectroscopically-observed GCs nicely sample the region in which the
intermediate-age GCs would be expected to lie.

Assuming that the higher S/N spectroscopic data and the improvements in the SSP models correctly reflect old ages for the entire GC sample,
we might conclude that the spectroscopic and photometric data are not entirely consistent. Figure 9 illustrates this point. The $V-I$
vs.~$V-K$ color distribution of the NGC 4365 GC system (panel 9a) is compared to that of NGC 3115 (panel 9b), an S0 galaxy with a
``normal'' GC color distribution, displaying the bimodality in $V-I$ typical of luminous galaxies. Spectroscopy of NGC 3115 GCs shows that
both the blue and red GCs are coeval and old (Kuntschner \etal~2002).  However, if all the NGC~4365 GCs are old as well, why do the color
distributions of these two galaxies look so different?

To try to address this question we have performed photometry for the central portion of the GC system using archival $g$ and $z$ HST/ACS
images obtained as part of the ACS Virgo Cluster Survey (C{\^ o}t{\' e} \etal~2004). Frames were processed through the standard pipeline,
including \emph{Multidrizzle} for image combining and rejection of cosmic rays. GC candidates were detected on 20 $\times$ 20 pixel
median-subtracted images. We performed photometry through a 5-pixel ($0.25\arcsec$) radius aperture, and corrected to a 10-pixel aperture
using corrections derived from bright objects of a range of sizes. These 10-pixel magnitudes were then corrected to a nominal infinite
aperture using the values in Sirianni \etal~(2004). These $g$ and $z$ magnitudes are on the AB system, using the zeropoints in Jordan
\etal~(2004). All photometry has been corrected for reddening using the maps of Schlegel \etal~(1998). The resulting color histogram is
shown in Figure 10, and in Figure 11 we show a similar histogram constructed for M87. In both histograms, density estimates using an
Epanechnikov kernel with a small bin size are overplotted. The apparent interpretation of the NGC 4365 histogram is that the GC colors are
trimodal in $g-z$, with peaks at $\sim 0.90$, 1.22, and 1.34 (with errors $\sim \pm \, 0.02-0.03$), although it was found to be unimodal in
$V-I$. However, the significance of the three peaks cannot be readily assessed. Experimentation with KMM tests (Ashman \etal~1993) on
simulated color distributions gave falsely high levels of significance for subpopulations which were not actually present, suggesting this
technique cannot be reliably applied to this multimodal data. We note that there have been previous claims in the literature for trimodal
GC subpopulations (e.g., Ostrov, Geisler, \& Forte 1993), but to our knowledge NGC 4365 is the first galaxy with supporting spectroscopic
evidence.

To determine whether any of these three observed peaks are consistent with values expected from the galaxy luminosity--GC color relation
for the blue or red subpopulations (Forbes, Brodie, \& Grillmair 1997; Larsen \etal~2001; Strader, Brodie, \& Forbes 2004) a conversion
between $V-I$ and $g-z$ is required. A rough conversion may be made with reference to M87, which has been observed in both colors. In M87
the $V-I$ peaks are at 0.96 and 1.18, while the $g-z$ peaks are 0.89 and 1.42. Thus a crude, preliminary conversion equation is: $V-I$ =
0.42 ($g-z$) + 0.59. Note that this equation assumes no age difference between the two populations; this is consistent with our
spectroscopic results. A corresponding linear fit to Maraston (2005) SSP models for broadband colors (using model parameters of 13 Gyr and
a Kroupa IMF) gives a similar relation with a slightly steeper slope: $V-I$ = 0.49 ($g-z$) + 0.45. Our empirical formula suggests that the
blue and red peaks in NGC 4365 occur at $V-I \sim 0.96$ and 1.15. The values predicted by the galaxy luminosity--GC color relations are
0.93 and 1.17, implying that these converted $g-z$ blue and red peaks do indeed place NGC 4365 approximately on the normal relation. A
plausible explanation is that the intermediate-metallicity peak filled in the gap between the two regular subpopulations in the original
$V-I$ data, resulting in an abnormally broad, apparently unimodal distribution. A more detailed analysis of the consistency of ACS and
WFPC2 photometry, as well as new NTT/SOFI $K$ photometry for NGC 4365 GCs, is forthcoming in Larsen \etal~(2005).

Eleven of the 21 GCs in our sample fall within the field of view of the ACS imaging, and thus have $g-z$ colors. To test whether the
trimodality observed in the $g-z$ color histogram is the same as that seen in the Lick index-index plots, we calculated the mean color for
each of the spectroscopic groups. Of the 11 spectroscopically observed GCs with ACS colors, two are blue, six orange, and three red. The
corresponding mean colors are $g-z = 0.96, 1.21$, and 1.32---quite similar to the peaks in the color histogram, especially when the small
sample size is considered. Both the spectroscopy and HST optical photometry point to a consistent picture of three old subpopulations of
GCs in NGC 4365. However, spectroscopy of a larger sample of GCs would be useful to test this conclusion.

Finally, we note that within the ACS field of view, the orange GCs appear to be very centrally concentrated, with few located outside a
radius of $\sim 1 \arcmin$. By contrast, the red and blue GCs extend to the edges of the ACS images. As is typical in massive galaxies, the
blues are less concentrated than the reds (Larsen \etal~2001).

\section{Discussion and Conclusions}

If these results are confirmed, NGC 4365 would be the first galaxy found to contain three old subpopulations of GCs. Is this elliptical a
pathological anomaly, or simply in the tail of a distribution of multimodal GC systems waiting to be discovered? Models which consider the
formation of GCs in the hierarchical merging paradigm of galaxy formation (e.g., Beasley \etal~2002) naturally predict substructure as a
straightforward consequence of the complex merging histories of massive early-type galaxies. While the Beasley \etal~paper (as well as the
bulk of other semi-analytic models from several years ago) predict significant \emph{age} substructure in the star formation histories of
these galaxies, more recent N-body/hydrodynamic models suggest early, rapid formation (e.g., Nagamine \etal~2004). This is consistent with
the old ages of all three GC subpopulations in NGC 4365. However, there remains the question of why this galaxy has three rather than two
clear subpopulations.

As mentioned above, the galaxy does not have properties notably different than other ellipticals of similar luminosity and environment. The
main body and counterrotating core of the galaxy both appear to be old and $\alpha$-enhanced (Davies \etal~2001), though this depends on
the SSP model used (see above). The $L_{B_\odot} / L_X$ ratio is typical for a galaxy of its luminosity (O`Sullivan \etal~2001). Its strong
triaxiality precludes the presence of a central black hole significantly more massive than that predicted by the $M-\sigma$ relation
(Statler \etal~2004). In short, the GC system of this galaxy is the most abnormal thing about it.

The fact that the metal-poor and metal-rich GC peaks lie at the values predicted for a galaxy of this luminosity suggests that the
intermediate-metallicity peak is the ``interloper'' in an otherwise normal GC system. The observed colors of this orange subpopulation
($g-z = 1.22 \sim V-I = 1.10$) correspond to galaxies with $M_V \sim -17$ in the red GC relation of Larsen \etal~(2001). Galaxies of this
luminosity have very few GCs and it is not yet clear whether, in general, they have a red peak at all (Lotz \etal~2004). Indeed, the
typical size of the \emph{entire} GC system of a single $M_V = -17$ galaxy is $\sim$ 20--30 GCs, compared with hundreds or more orange GCs
in NGC 4365. Thus, accretion (C{\^ o}t{\' e} \etal~2002) is an unlikely explanation for this extra subpopulation.

The large number of these intermediate-metallicity clusters implies an unusual, quite important star formation event early in the history
of the galaxy, which might also have formed a significant fraction of the field star population. To estimate the amount of gas involved in
such an event we adopt a typical GC formation efficiency of 0.25\% (McLaughlin 1999) and a typical GC mass of $3 \times 10^5 M_\odot$.
Roughly $\sim 1.2 \times 10^8 M_\odot$ of gas are needed to form one new GC (Harris 2003). A galaxy of the luminosity of NGC 4365 typically
has similar numbers of metal-poor and metal-rich GCs (Forbes \& Forte 2001); our color histogram suggests that perhaps 40\% of the non-blue
GCs (and thus $\sim 20\%$ of the total GC population) are of intermediate-metallicity. For a total GC population of $\sim 2500$ (Ashman \&
Zepf 1998), this would imply $\sim 500$ orange clusters are formed. The corresponding gas mass is large: $6 \times 10^{10} M_\odot$.

However, if the field stars are primarily old, the intermediate-metallicity population would probably not be noticeable in the integrated
spectrum of the galaxy itself. The luminosity-weighted metallicity of the galaxy would appear slightly lower than otherwise, but far less
than would be needed to make the galaxy an outlier in the metallicity hyperplane (Trager \etal~2000). A color-magnitude diagram of the red
giant branch of the galaxy itself (possible with a 30-m telescope) will probably be needed to quantitatively understand the relative
importance of this star formation event in building up the galaxy's stellar mass.

In conclusion, we find evidence for three old subpopulations of GCs in NGC 4365. No metal-rich intermediate-age GCs are apparent, though
the size (21 GCs) and spatial distribution (few within the central 4 kpc) of our sample are caveats to our results. The highly multiplexing
spectrographs now on 8--10-m class telescopes (e.g., Keck/DEIMOS, VLT/VIMOS) could permit one to obtain spectra of 100-150 GCs in a single
slitmask. However, the strong central concentration of GC systems (with typical surface densities of $\sim r^{-1.5}$ to $r^{-2}$) will
likely limit such studies to the most populous systems for the foreseeable future.

Our results also indicate that it is difficult to identify intermediate-age GCs solely from broadband photometry, even if a NIR band is
used. It will be interesting to see if evidence for significant intermediate-age subpopulations in other galaxies, e.g., NGC 5846 (Hempel
\etal~2003), is supported by spectroscopic follow-up.

\acknowledgments

We thank the anonymous referee for detailed comments on the paper. We are grateful to Thomas Puzia for providing us with his photometry and
to Claudia Maraston and Daniel Thomas for early access to their models, as well as helpful discussions. Ricardo Schiavon provided useful
insights. The work was supported by NSF grant number AST-0206139. J.~S.~acknowledges support from an NSF Graduate Research Fellowship.
A.~J.~C.~acknowledges financial support from a UCM Fundaci\'on del Amo Fellowship. Spectroscopic data presented herein were obtained at the
W.~M.~Keck Observatory, which is operated as a scientific partnership among the California Institute of Technology, the University of
California, and the National Aeronautics and Space Administration. The Observatory was made possible by the generous financial support of
the W.~M.~Keck Foundation. The authors wish to recognize and acknowledge the very significant cultural role and reverence that the summit
of Mauna Kea has always had within the indigenous Hawaiian community. We are most fortunate to have the opportunity to conduct observations
from this mountain.

{}

\clearpage

\begin{deluxetable}{crrrrr}
\tabletypesize{\scriptsize}
\tablecaption{Globular cluster candidates in NGC~4365.\label{tbl-1}}
\tablewidth{0pt}
\tablehead{
\colhead{ID}          & 
\colhead{R.A.}        & 
\colhead{Dec.}       &
\colhead{V}       &
\colhead{V$_{helio}$}   &
\colhead{GC?}\\
\colhead{} &
\colhead{(J2000.0)} &
\colhead{(J2000.0)} &
\colhead{(mag)} &
\colhead{(km/s)} &
\colhead{}}
\startdata

1      & 12:24:23.9 & 7:16:36 & 21.2 &  752 $\pm$ 14 & GC \\		
2     & 12:24:24.1 & 7:17:21 & 22.0 & 1732 $\pm$ 14 & GC \\		
3     & 12:24:23.7 & 7:17:36 & 21.7 &  635 $\pm$ 15 & GC \\
4     & 12:24:25.8 & 7:17:33 & 21.9 & 1030 $\pm$ 14 & GC \\
5     & 12:24:26.8 & 7:17:37 & 21.5 & 1226 $\pm$ 11 & GC \\
6     & 12:24:25.4 & 7:17:56 & 22.3 & 1418 $\pm$ 14 & GC \\		
7     & 12:24:27.0 & 7:17:58 & 22.0 &  919 $\pm$ 12 & GC \\		
8   & 12:24:25.3 & 7:18:27 & 21.2 & 1592 $\pm$ 12 & GC \\
9    & 12:24:27.2 & 7:18:22 & 21.0 & 1252 $\pm$ 13 & GC \\
10   & 12:24:27.3 & 7:18:35 & 18.4 &  111 $\pm$ 16 & star \\
11   & 12:24:29.3 & 7:18:40 & 22.0 & 1326 $\pm$ 11 & GC \\
12    & 12:24:30.5 & 7:18:51 & 20.7 & 1329 $\pm$  9 & GC \\
13 & 12:24:27.4 & 7:19:44 & 21.9 &  496 $\pm$ 14 & GC \\
14   & 12:24:31.0 & 7:19:28 & 21.4 & 1563 $\pm$ 10 & GC \\
15   & 12:24:31.0 & 7:19:54 & 21.7 & 1097 $\pm$ 10 & GC \\
16  & 12:24:30.1 & 7:20:13 & 21.0 &  983 $\pm$ 15 & GC \\
17 & 12:24:30.3 & 7:20:23 & 21.5 &  898 $\pm$ 14 & GC \\
18   & 12:24:31.8 & 7:20:32 & 21.4 & 1206 $\pm$  9 & GC \\
19   & 12:24:33.0 & 7:20:53 & 21.5 & 1246 $\pm$ 13 & GC \\
20   & 12:24:33.4 & 7:21:02 & 21.2 & 1501 $\pm$ 13 & GC \\		
21    & 12:24:34.9 & 7:21:17 & 22.5 &  986 $\pm$  9 & GC \\		
22  & 12:24:33.2 & 7:21:38 & 22.3 &  985 $\pm$ 12 & GC \\		
23    & 12:24:33.6 & 7:22:03 & 22.0 &  980 $\pm$ 16 & GC \\		
24   & 12:24:35.1 & 7:22:16 & 22.3 &22230 $\pm$ 34 & galaxy  \\

\enddata
\end{deluxetable}

\begin{deluxetable}{ccccccccccccccccccccc}
\rotate
\setlength{\tabcolsep}{0.017in}
\tabletypesize{\footnotesize}
\tablecaption{Lick indices of globular clusters.\label{tbl-2}}
\tablewidth{0pt}
\tablehead{
\colhead{ID} & 
\colhead{H$\delta_A$} & 
\colhead{H$\delta_F$} & 
\colhead{CN$_1$} & 
\colhead{CN$_2$} & 
\colhead{Ca4227} & 
\colhead{G4300} & 
\colhead{H$\gamma_A$} & 
\colhead{H$\gamma_F$} & 
\colhead{Fe4383} & 
\colhead{Ca4455} & 
\colhead{Fe4531} & 
\colhead{Fe4668} & 
\colhead{H$\beta$} & 
\colhead{Fe5015} & 
\colhead{Mg1} & 
\colhead{Mg2} & 
\colhead{Mg$b$} & 
\colhead{Fe5270} & 
\colhead{Fe5335} & 
\colhead{Fe5406}\\
\colhead{} &
\colhead{(\AA)} &
\colhead{(\AA)} &
\colhead{(mag)} &
\colhead{(mag)} &
\colhead{(\AA)} &
\colhead{(\AA)} &
\colhead{(\AA)} &
\colhead{(\AA)} &
\colhead{(\AA)} &
\colhead{(\AA)} &
\colhead{(\AA)} &
\colhead{(\AA)} &
\colhead{(\AA)} &
\colhead{(\AA)} &
\colhead{(mag)} &
\colhead{(mag)} &
\colhead{(\AA)} &
\colhead{(\AA)} &
\colhead{(\AA)} &
\colhead{(\AA)}}
\startdata

1 &   1.47 &  1.80 &  0.003 &  0.031 & 0.29 & 3.10 & -1.20 &  1.06 & 2.02 & 0.71 & 2.12 & 0.85 & 2.64 & 2.37 & 0.042 & 0.102 & 1.53 & 1.39 & 1.48 & 1.01\\
2 &   1.46 &  1.15 & -0.012 &  0.015 & 0.66 & 3.49 & -1.27 &  0.44 & 1.24 & 0.76 & 1.75 & 1.77 & 2.29 & 2.37 & 0.039 & 0.105 & 1.86 & 1.67 & 1.05 & 1.06\\
3 &  -0.08 &  0.32 &  0.037 &  0.048 & 0.47 & 3.89 & -2.93 &  0.55 & 4.56 & 0.40 & 2.31 & 2.09 & 1.76 & 2.54 & 0.068 & 0.162 & 2.59 & 1.60 & 1.55 & 1.08\\
5 &  -1.94 & -0.11 &  0.076 &  0.111 & 0.90 & 5.58 & -5.16 & -0.71 & 4.38 & 1.04 & 2.42 & 2.89 & 1.82 & 5.82 & 0.068 & 0.199 & 3.29 & 3.00 & 1.63 & 1.24\\
6 &   1.52 &  1.16 & -0.053 & -0.037 & 0.15 & 2.59 & -0.13 &  1.29 & 1.90 & 0.22 & 1.44 & 0.02 & 2.56 & 2.97 & 0.030 & 0.101 & 1.54 & 1.43 & 0.46 & 0.38\\
7 &   1.49 &  1.38 & -0.031 & -0.008 & 0.44 & 3.66 & -1.57 &  0.75 & 1.00 & 0.06 & 1.00 & 0.75 & 1.86 & 2.81 & 0.059 & 0.108 & 1.72 & 1.23 & 0.74 & 0.15\\
8 &  -0.13 &  0.95 &  0.048 &  0.087 & 1.11 & 4.44 & -3.66 & -0.20 & 3.62 & 0.80 & 2.86 & 2.82 & 1.88 & 4.46 & 0.071 & 0.190 & 3.29 & 2.03 & 1.39 & 1.38\\
9 &  -0.65 &  1.12 &  0.075 &  0.117 & 0.69 & 4.21 & -3.34 & -0.05 & 3.28 & 0.91 & 2.23 & 2.04 & 1.50 & 3.26 & 0.083 & 0.188 & 2.99 & 2.49 & 1.82 & 1.27\\
11 & -0.86 & -0.81 &  0.063 &  0.068 & 0.71 & 5.26 & -3.95 &  0.40 & 4.08 & 1.31 & 3.44 & 3.31 & 2.15 & 6.81 & 0.076 & 0.206 & 3.15 & 2.45 & 1.00 & 0.39\\
12 & -0.30 &  0.80 &  0.046 &  0.081 & 0.80 & 4.22 & -3.24 &  0.02 & 3.15 & 1.06 & 3.08 & 2.48 & 2.23 & 3.95 & 0.051 & 0.172 & 3.10 & 1.87 & 1.65 & 1.28\\
13 & -0.21 &  0.98 &  0.068 &  0.098 & 1.26 & 3.89 & -3.15 &  0.10 & 2.53 & 1.41 & 3.11 & 5.37 & 1.08 & 5.22 & 0.087 & 0.214 & 3.47 & 2.10 & 1.49 & 1.35\\
14 & -0.24 &  0.80 &  0.032 &  0.070 & 0.87 & 4.85 & -3.62 & -0.02 & 3.51 & 0.97 & 2.15 & 3.35 & 2.02 & 4.30 & 0.068 & 0.180 & 3.02 & 1.68 & 1.65 & 0.80\\
15 & -0.61 &  0.73 &  0.066 &  0.100 & 0.92 & 4.32 & -3.93 & -0.57 & 3.46 & 0.83 & 3.27 & 3.62 & 1.67 & 5.28 & 0.082 & 0.226 & 3.78 & 2.74 & 1.61 & 1.33\\
16 & -2.05 & -0.08 &  0.136 &  0.177 & 1.06 & 5.07 & -5.48 & -1.24 & 4.80 & 1.32 & 3.35 & 4.24 & 1.45 & 4.82 & 0.107 & 0.258 & 4.41 & 2.78 & 2.11 & 1.66\\
17 &  0.10 &  0.88 &  0.061 &  0.106 & 0.45 & 3.91 & -2.72 &  0.10 & 2.98 & 0.77 & 2.56 & 2.80 & 1.85 & 3.44 & 0.061 & 0.159 & 2.29 & 1.97 & 1.67 & 1.15\\
18 & -1.39 &  0.57 &  0.106 &  0.143 & 0.68 & 4.54 & -4.31 & -0.48 & 4.01 & 1.05 & 2.67 & 3.29 & 2.00 & 4.84 & 0.090 & 0.219 & 3.69 & 2.59 & 1.86 & 1.35\\
19 & -0.18 &  0.57 &  0.047 &  0.075 & 0.69 & 4.58 & -4.17 & -1.08 & 1.74 & 0.75 & 3.23 & 1.56 & 1.76 & 4.03 & 0.059 & 0.180 & 3.14 & 1.81 & 1.81 & 1.20\\
20 &  0.89 &  1.28 & -0.016 &  0.011 & 0.66 & 4.16 & -2.44 &  0.43 & 2.21 & 0.77 & 2.25 & 1.84 & 2.06 & 3.31 & 0.041 & 0.116 & 1.83 & 1.26 & 1.32 & 0.97\\
21 &  2.15 &  1.77 & -0.003 &  0.027 & 0.68 & 4.03 & -2.04 &  0.77 & 3.88 & 1.61 & 1.01 & 0.87 & 2.30 & 4.49 & 0.049 & 0.116 & 1.87 & 1.27 & 2.45 & 0.78\\
22 &  2.04 &  2.15 & -0.005 &  0.032 & 0.22 & 3.42 & -2.76 &  0.02 & 2.97 & 0.17 & 1.82 & 1.95 & 2.44 & 2.56 & 0.025 & 0.124 & 3.17 & 0.63 & 1.41 & 1.65\\
23 &  1.31 &  1.66 & -0.016 &  0.025 & 0.25 & 3.14 &  0.04 &  1.62 & 0.98 & 0.50 & 1.74 & 1.19 & 1.71 & 3.04 & 0.031 & 0.104 & 2.03 & 1.26 & 1.69 & 0.48\\

\enddata
\end{deluxetable}

\begin{deluxetable}{ccccccccccccccccccccc}
\setlength{\tabcolsep}{0.02in}
\rotate
\tabletypesize{\footnotesize} 
\tablecaption{Lick index errors.\label{tbl-2a}}
\tablewidth{0pt}
\tablehead{
\colhead{ID} & 
\colhead{H$\delta_A$} & 
\colhead{H$\delta_F$} & 
\colhead{CN$_1$} & 
\colhead{CN$_2$} & 
\colhead{Ca4227} & 
\colhead{G4300} & 
\colhead{H$\gamma_A$} & 
\colhead{H$\gamma_F$} & 
\colhead{Fe4383} & 
\colhead{Ca4455} & 
\colhead{Fe4531} & 
\colhead{Fe4668} & 
\colhead{H$\beta$} & 
\colhead{Fe5015} & 
\colhead{Mg1} & 
\colhead{Mg2} & 
\colhead{Mg$b$} & 
\colhead{Fe5270} & 
\colhead{Fe5335} & 
\colhead{Fe5406}\\
\colhead{} &
\colhead{(\AA)} &
\colhead{(\AA)} &
\colhead{(mag)} &
\colhead{(mag)} &
\colhead{(\AA)} &
\colhead{(\AA)} &
\colhead{(\AA)} &
\colhead{(\AA)} &
\colhead{(\AA)} &
\colhead{(\AA)} &
\colhead{(\AA)} &
\colhead{(\AA)} &
\colhead{(\AA)} &
\colhead{(\AA)} &
\colhead{(mag)} &
\colhead{(mag)} &
\colhead{(\AA)} &
\colhead{(\AA)} &
\colhead{(\AA)} &
\colhead{(\AA)}}
\startdata

1 &  0.35 & 0.19 & 0.008 & 0.008 & 0.17 & 0.31 & 0.27 & 0.16 & 0.40 & 0.19 & 0.26 & 0.47 & 0.18 & 0.50 & 0.006 & 0.009 & 0.17 & 0.25 & 0.27 & 0.28\\
2 &  0.48 & 0.26 & 0.012 & 0.012 & 0.16 & 0.47 & 0.71 & 0.32 & 0.64 & 0.16 & 0.40 & 0.76 & 0.46 & 0.62 & 0.010 & 0.013 & 0.26 & 0.45 & 0.38 & 0.40\\
3 &  0.54 & 0.39 & 0.019 & 0.019 & 0.14 & 0.26 & 0.46 & 0.27 & 0.68 & 0.24 & 0.33 & 0.68 & 0.25 & 0.50 & 0.009 & 0.009 & 0.19 & 0.49 & 0.33 & 0.19\\
5 &  0.40 & 0.33 & 0.009 & 0.010 & 0.18 & 0.43 & 0.56 & 0.29 & 0.60 & 0.23 & 0.31 & 0.54 & 0.22 & 0.69 & 0.007 & 0.010 & 0.35 & 0.34 & 0.39 & 0.36\\
6 &  0.57 & 0.44 & 0.016 & 0.019 & 0.31 & 0.60 & 0.49 & 0.36 & 0.68 & 0.48 & 0.66 & 1.16 & 0.29 & 0.89 & 0.012 & 0.020 & 0.48 & 0.60 & 0.43 & 0.40\\
7 &  0.58 & 0.28 & 0.012 & 0.016 & 0.20 & 0.48 & 0.55 & 0.31 & 0.71 & 0.27 & 0.55 & 1.07 & 0.31 & 1.27 & 0.008 & 0.014 & 0.48 & 0.55 & 0.51 & 0.67\\
8 &  0.41 & 0.24 & 0.011 & 0.010 & 0.17 & 0.35 & 0.41 & 0.25 & 0.42 & 0.28 & 0.33 & 0.53 & 0.25 & 0.38 & 0.008 & 0.009 & 0.33 & 0.31 & 0.40 & 0.18\\
9 &  0.35 & 0.21 & 0.007 & 0.011 & 0.15 & 0.25 & 0.33 & 0.22 & 0.36 & 0.18 & 0.32 & 0.58 & 0.23 & 0.51 & 0.005 & 0.009 & 0.28 & 0.32 & 0.38 & 0.13\\
11 & 0.92 & 0.45 & 0.025 & 0.025 & 0.34 & 0.71 & 0.93 & 0.47 & 0.68 & 0.52 & 0.54 & 1.18 & 0.47 & 1.54 & 0.012 & 0.022 & 0.72 & 0.92 & 0.65 & 0.55\\
12 & 0.22 & 0.14 & 0.004 & 0.005 & 0.12 & 0.18 & 0.29 & 0.17 & 0.21 & 0.12 & 0.45 & 0.37 & 0.11 & 0.42 & 0.005 & 0.005 & 0.17 & 0.20 & 0.13 & 0.20\\
13 & 0.57 & 0.34 & 0.019 & 0.022 & 0.33 & 0.56 & 0.72 & 0.44 & 0.72 & 0.48 & 0.47 & 0.96 & 0.30 & 0.79 & 0.014 & 0.013 & 0.47 & 0.29 & 0.40 & 0.42\\
14 & 0.28 & 0.30 & 0.015 & 0.015 & 0.15 & 0.41 & 0.36 & 0.26 & 0.37 & 0.22 & 0.24 & 0.65 & 0.31 & 0.44 & 0.007 & 0.006 & 0.16 & 0.30 & 0.49 & 0.23\\
15 & 0.46 & 0.36 & 0.013 & 0.013 & 0.18 & 0.31 & 0.45 & 0.37 & 0.65 & 0.24 & 0.31 & 0.72 & 0.19 & 0.79 & 0.005 & 0.011 & 0.29 & 0.43 & 0.41 & 0.24\\
16 & 0.26 & 0.17 & 0.011 & 0.012 & 0.09 & 0.20 & 0.26 & 0.13 & 0.27 & 0.13 & 0.19 & 0.36 & 0.13 & 0.25 & 0.004 & 0.005 & 0.14 & 0.16 & 0.27 & 0.12\\
17 & 0.32 & 0.22 & 0.008 & 0.010 & 0.11 & 0.27 & 0.27 & 0.16 & 0.43 & 0.14 & 0.29 & 0.32 & 0.25 & 0.46 & 0.007 & 0.007 & 0.30 & 0.23 & 0.32 & 0.21\\
18 & 0.31 & 0.24 & 0.015 & 0.016 & 0.20 & 0.28 & 0.31 & 0.18 & 0.27 & 0.14 & 0.25 & 0.66 & 0.19 & 0.37 & 0.006 & 0.007 & 0.19 & 0.17 & 0.23 & 0.17\\
19 & 0.32 & 0.21 & 0.009 & 0.012 & 0.15 & 0.31 & 0.42 & 0.18 & 0.48 & 0.16 & 0.52 & 0.56 & 0.24 & 0.66 & 0.009 & 0.008 & 0.19 & 0.29 & 0.37 & 0.21\\
20 & 0.28 & 0.14 & 0.009 & 0.010 & 0.11 & 0.28 & 0.23 & 0.15 & 0.27 & 0.18 & 0.29 & 0.38 & 0.12 & 0.39 & 0.005 & 0.006 & 0.15 & 0.12 & 0.19 & 0.13\\
21 & 1.01 & 0.78 & 0.025 & 0.024 & 0.33 & 0.95 & 0.86 & 0.68 & 0.85 & 0.26 & 0.72 & 1.52 & 0.40 & 1.19 & 0.008 & 0.014 & 0.46 & 0.68 & 0.51 & 0.43\\
22 & 0.69 & 0.55 & 0.022 & 0.028 & 0.26 & 0.58 & 0.86 & 0.56 & 0.96 & 0.49 & 1.04 & 1.56 & 0.38 & 0.99 & 0.015 & 0.018 & 0.49 & 0.46 & 0.39 & 0.44\\
23 & 0.37 & 0.23 & 0.012 & 0.013 & 0.13 & 0.33 & 0.39 & 0.29 & 0.67 & 0.28 & 0.48 & 0.54 & 0.34 & 0.76 & 0.009 & 0.013 & 0.35 & 0.28 & 0.37 & 0.26\\

\enddata
\end{deluxetable}

\clearpage

\begin{deluxetable}{lr}
\tabletypesize{\scriptsize}
\tablecaption{Corrections to the Lick System \label{tbl-3a}}
\tablewidth{0pt}
\tablehead{
\colhead{Index}          &
\colhead{Offset (Lick/IDS $-$ This Paper)}}
\startdata
H$\delta_A$ &  0.01 $\pm$  0.20 \AA\\
H$\delta_F$ &  $-$0.07 $\pm$  0.24 \AA\\
CN$_1$ &  0.001 $\pm$  0.007 mag\\
CN$_2$ &  0.008 $\pm$  0.008 mag\\
Ca4227 &  0.13 $\pm$  0.13 \AA\\
G4300  &  $-$0.13 $\pm$  0.21 \AA\\
H$\gamma_A$ &  0.18 $\pm$  0.23 \AA\\
H$\gamma_F$ &  0.01 $\pm$  0.04 \AA\\
Fe4383 &  0.52 $\pm$  0.35 \AA\\
Ca4455 &  0.09 $\pm$  0.26 \AA\\
Fe4531 &  0.30 $\pm$  0.24 \AA\\
C$_2$4668 &  0.13 $\pm$  0.18 \AA\\
H$\beta$  &  $-$0.02 $\pm$  0.02 \AA\\
Fe5015 &  0.16 $\pm$  0.42 \AA\\
Mg$_1$ &  0.006 $\pm$  0.006 mag\\
Mg$_2$ &  0.008 $\pm$  0.005 mag\\
Mg{\it b}  & 0.08 $\pm$  0.15 \AA\\
Fe5270 &  0.11 $\pm$  0.16 \AA\\
Fe5335 &  0.06 $\pm$  0.28 \AA\\
Fe5406 &  0.03 $\pm$  0.15 \AA\\
\enddata

\end{deluxetable}

\clearpage

\begin{deluxetable}{cccc}
\tabletypesize{\footnotesize}
\tablecaption{Spectroscopic metallicity estimates and subpopulations\label{tbl-3}}
\tablewidth{0pt}
\tablehead{
\colhead{ID} &
\colhead{PCA [m/H]} &
\colhead{$\chi^2$ [m/H]} &
\colhead{Subpopulation}\\
\colhead{} &
\colhead{(dex)} &
\colhead{(dex)} &
\colhead{}}
\startdata

1   &  $-1.10$  &  $-1.08$  	&  Blue\\
2   &  $-1.01$  &  $-0.95$	&  Blue\\
3   &  $-0.61$  &  $-0.68$	&  Orange\\
5   &  $-0.12$  &  $-0.23$	&  Red\\
6   &  $-1.34$  &  $-1.18$	&  Blue\\
7   &  $-1.13$  &  $-1.03$	&  Blue\\
8   &  $-0.36$  &  $-0.28$	&  Orange\\
9   &  $-0.32$  &  $-0.53$	&  Orange\\
11  &  $-0.43$  &  $-0.23$	&  Orange\\
12  &  $-0.52$  &  $-0.40$	&  Orange\\
13  &  $-0.25$  &  $-0.25$	&  Orange\\
14  &  $-0.46$  &  $-0.43$	&  Orange\\
15  &  $-0.21$  &  $-0.18$	&  Red\\
16  &  $0.18$   &  $0.03$	&  Red\\
17  &  $-0.60$  &  $-0.48$	&  Orange\\
18  &  $-0.20$  &  $-0.13$	&  Red\\
19  &  $-0.49$  &  $-0.43$	&  Orange\\
20  &  $-0.90$  &  $-0.85$	&  Blue\\
21  &  $-0.73$  &  $-0.85$	&  Blue\\
22  &  $-1.01$  &  $-0.85$	&  Blue\\
23  &  $-1.04$  &  $-1.08$	&  Blue\\

\enddata
\end{deluxetable}

\clearpage

\begin{figure}
\plotone{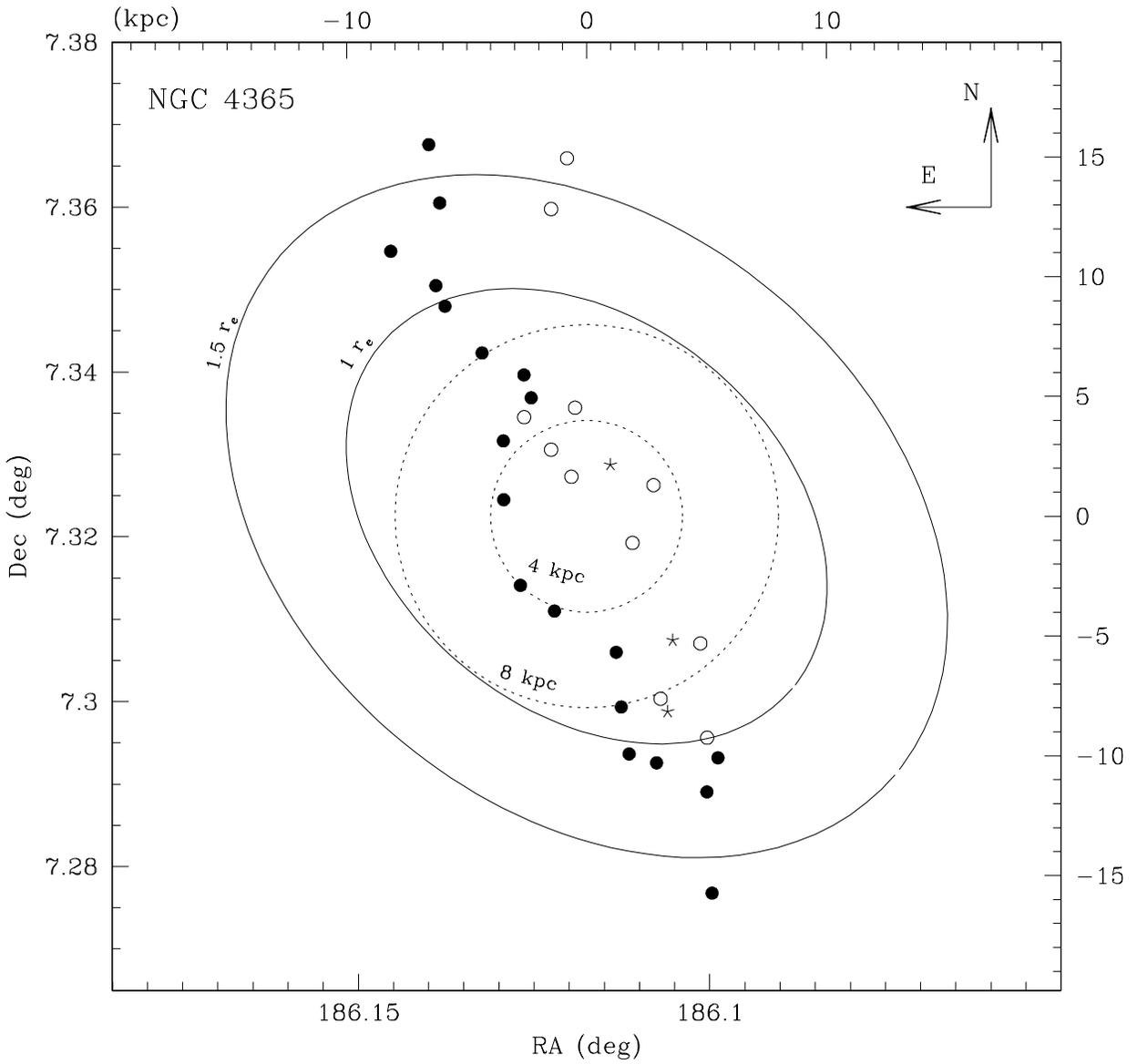}
\caption{Spatial distribution for NGC 4365 GCs. The filled circles are those GCs with spectra from this work; the open circles are those from Paper I. Stars 
are overlapping GCs from the two samples. Dotted circles represent radii of 4 and 8 kpc, while the solid ellipses represent the approximate isophotal contours
of the stellar light at 1 and 1.5 r$_e$.}
\end{figure}
\clearpage

\begin{figure}
\plotone{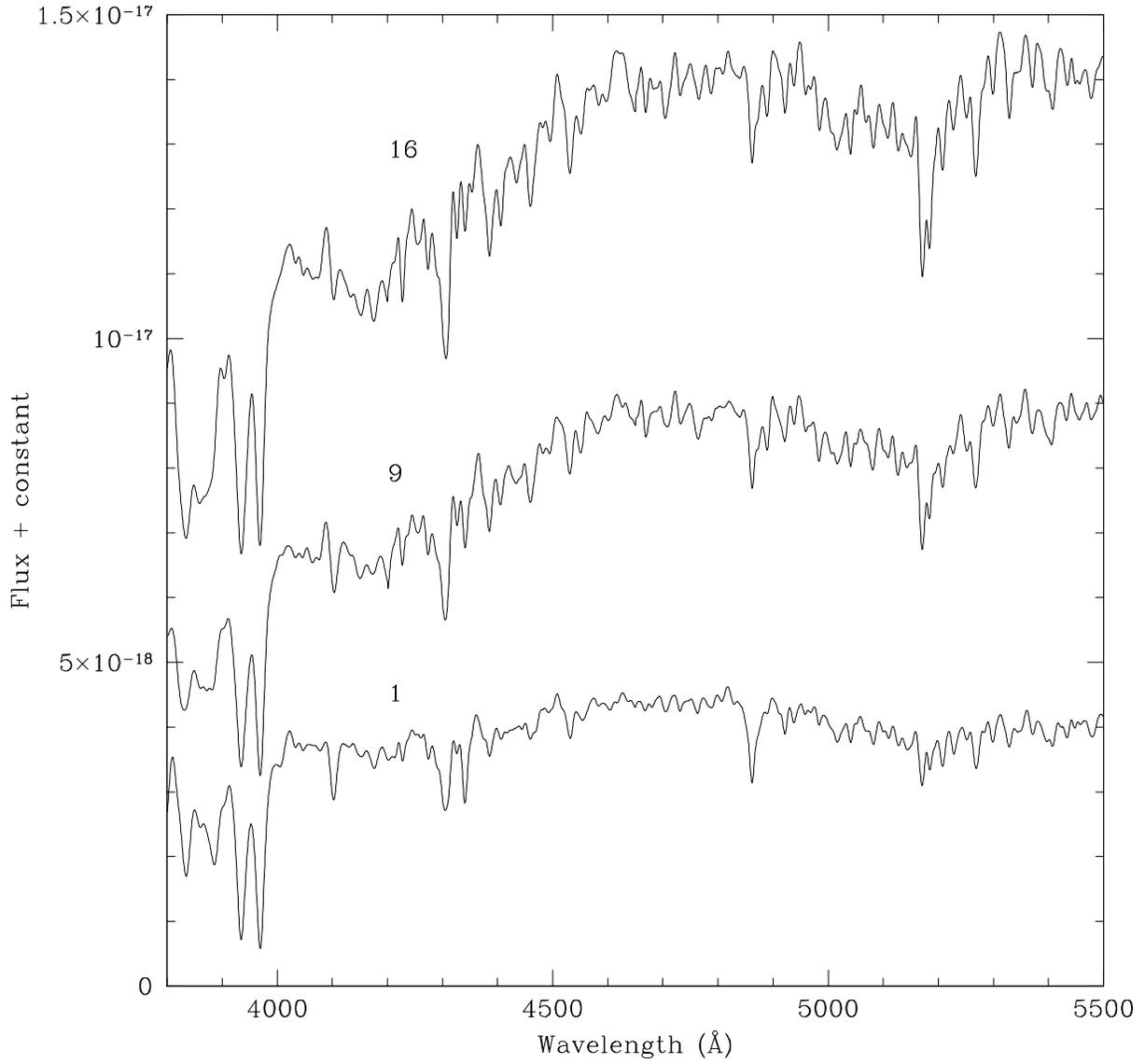}
\caption{Spectra of GCs 1 (metal-poor), 9 (intermediate-metallicity), and 16 (metal-rich), smoothed to the Lick resolution. See text for details on the GC subpopulations.}
\end{figure}  
\clearpage

\begin{figure}
\plotone{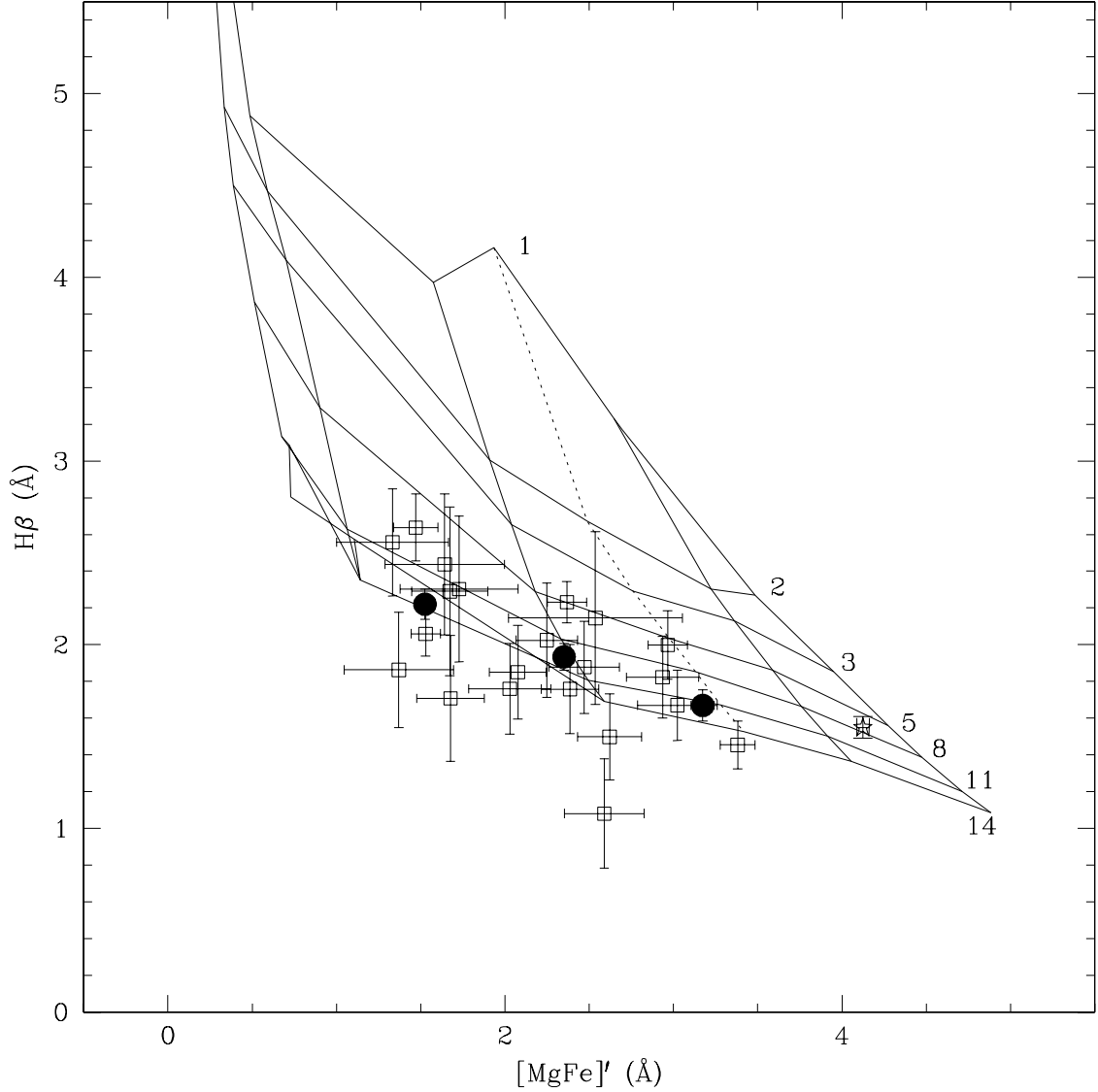}
\caption{H$\beta$-[MgFe]$\arcmin$ diagnostic diagrams for the globular clusters (squares) and central r$_e$/8 integrated light (star) of the elliptical
galaxy NGC 4365. The filled circles are weighted means for the three subpopulations (see text). Overlaid are SSP model grids from Thomas, Maraston, \& Korn 
(2004). From top to bottom, the isochrone ages are 1, 2, 3, 5, 8, 11, and 14 Gyr. From left to right, the isosiders (constant metallicity lines) are [$Z$/H] = 
$-2.25, -1.33, -0.33, 0$ (dotted), 0.33, and 0.67. All of the clusters appear to be old.}
\end{figure}
\clearpage

\begin{figure}
\plotone{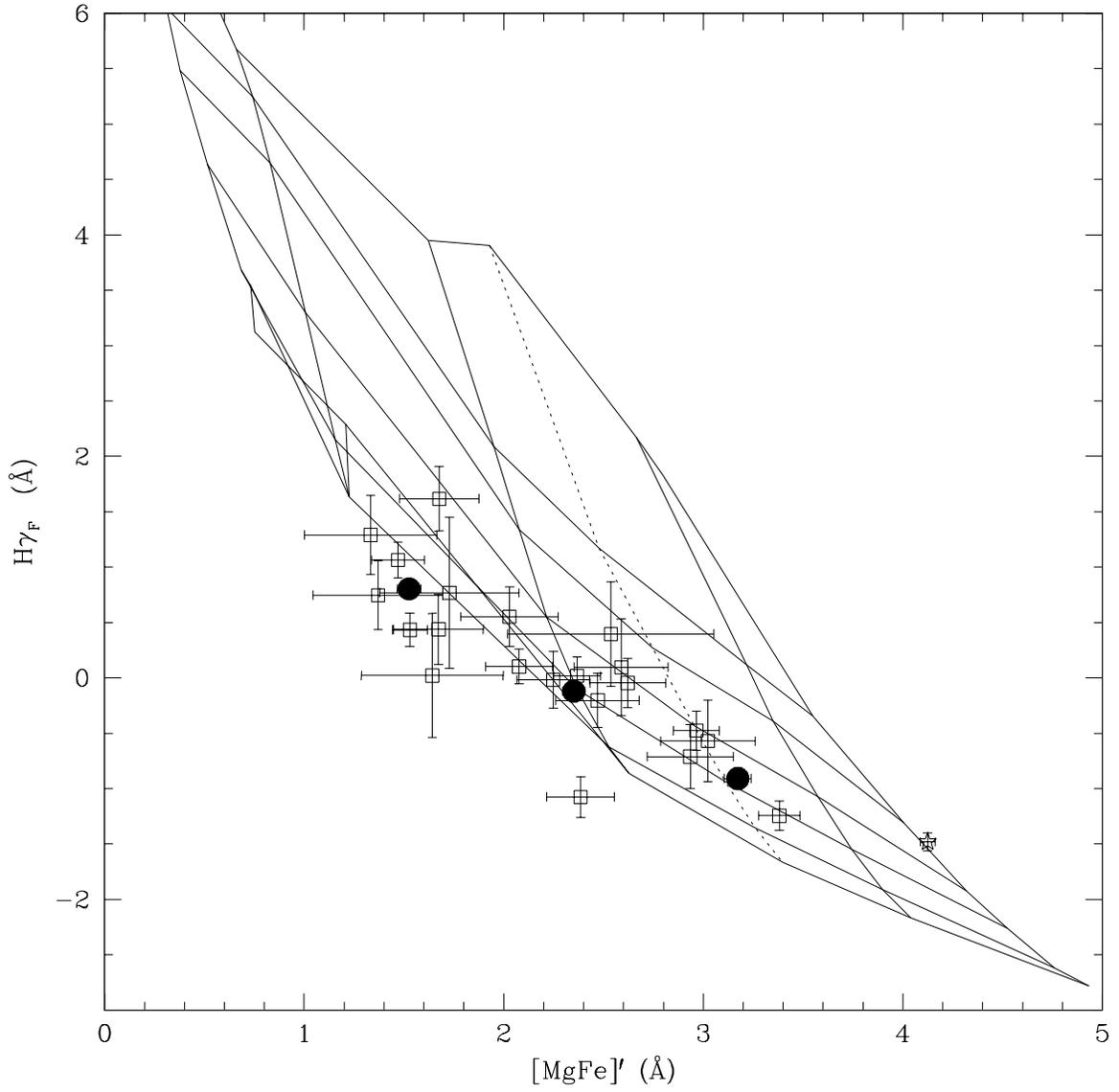}
\caption{H$\gamma_F$-[MgFe]$\arcmin$ diagram. Symbols and grid identities are as in Figure 2.}
\end{figure}
\clearpage

\begin{figure}
\plotone{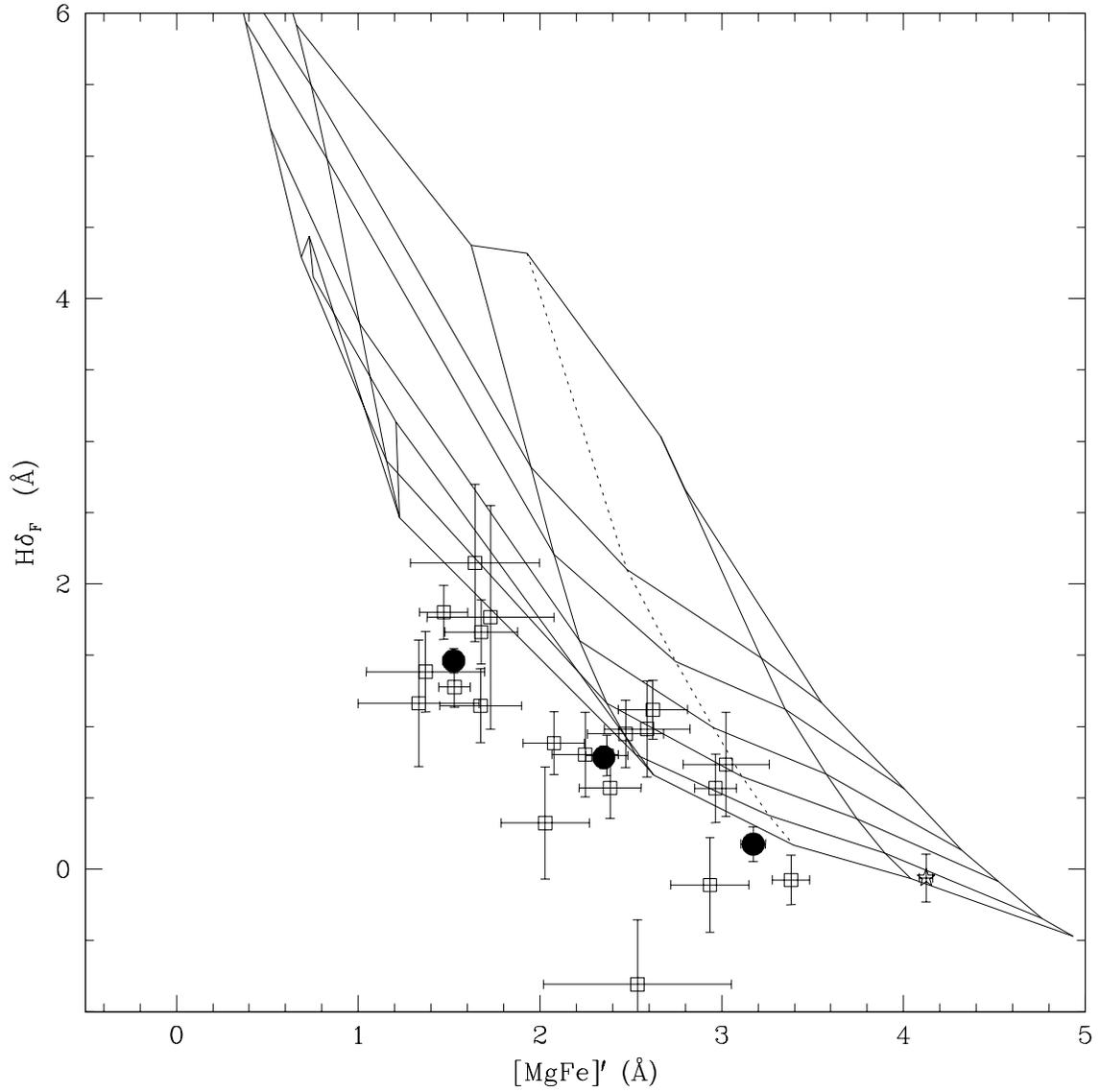}
\caption{H$\delta_F$-[MgFe]$\arcmin$ diagram. Symbols and grid identities are as in Figure 2. Three old subpopulations of globular clusters are apparent.}
\end{figure}
\clearpage  

\begin{figure}
\plotone{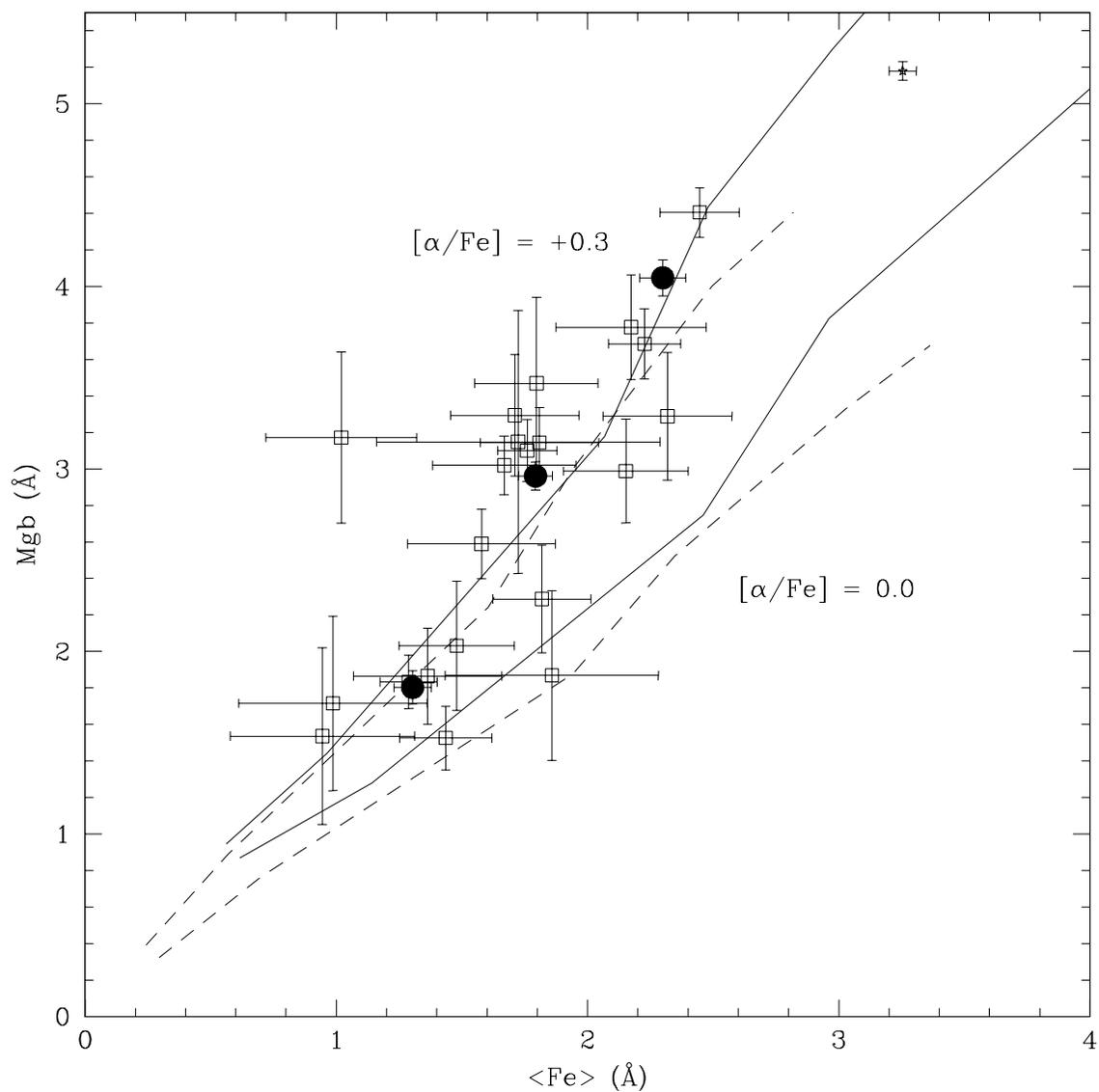}
\caption{Mg$b$-$<$Fe$>$ diagram. Symbols are as in Figure 2. 14 Gyr (solid) and 3 Gyr (dashed) isochrones for the labeled [$\alpha$/Fe] ratios (from TMK04) are 
shown. The orange and red GCs are $\alpha$-enhanced to values typical of Galactic GCs.}
\end{figure} 
\clearpage

\begin{figure}
\plotone{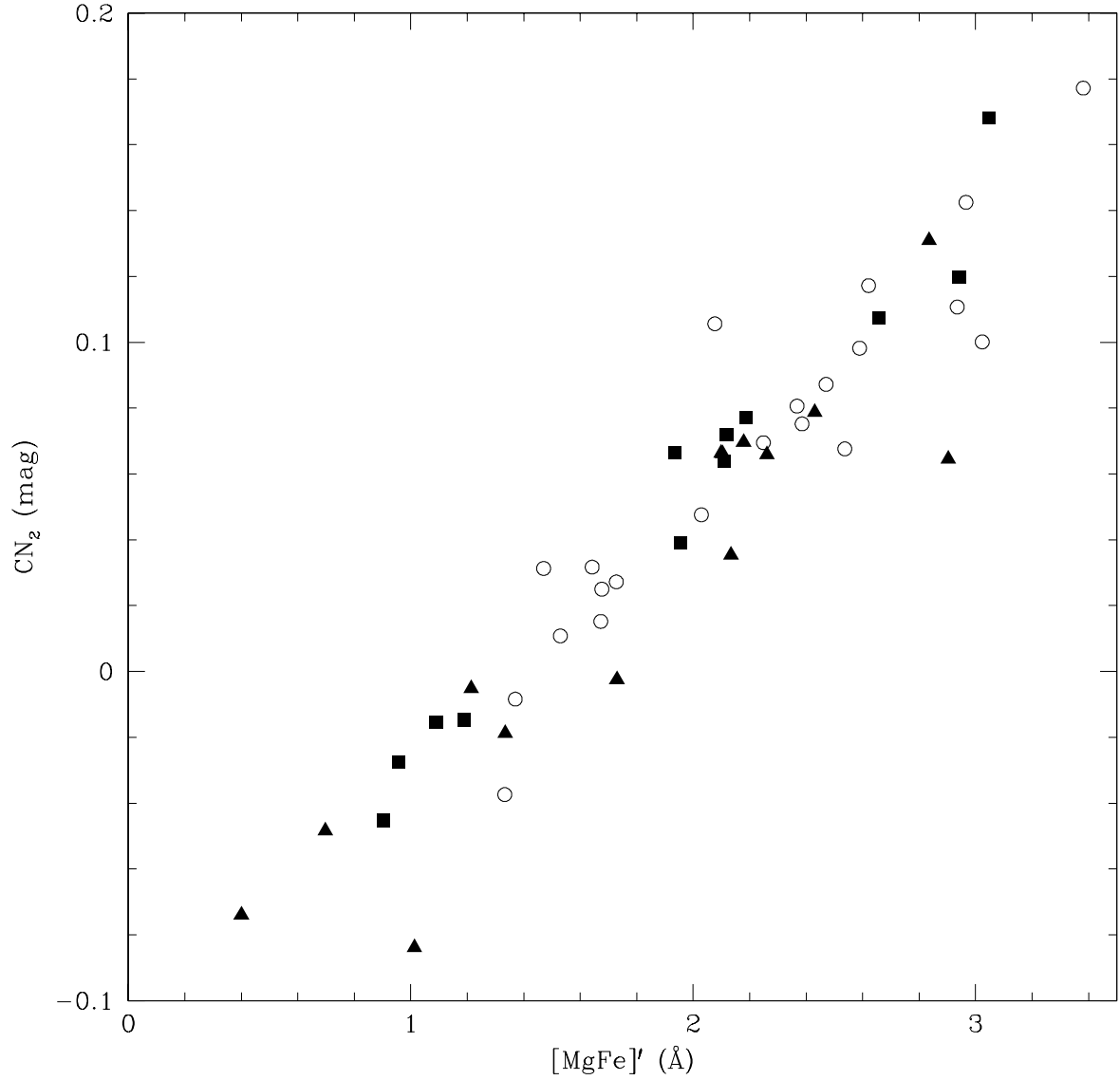}
\caption{CN$_2$-[MgFe]$\arcmin$ diagram. Plotted are our sample GCs (open circles) and Galactic GCs from Puzia \etal~(2003) and Gregg (1995), filled squares 
and triangles respectively. The NGC 4365 GCs do not appear to be enhanced in CN with respect to Galactic GCs; however, the sensitivity of CN$_2$ to CN variations is weak 
(see text).}
\end{figure}
\clearpage  

\begin{figure}
\plotone{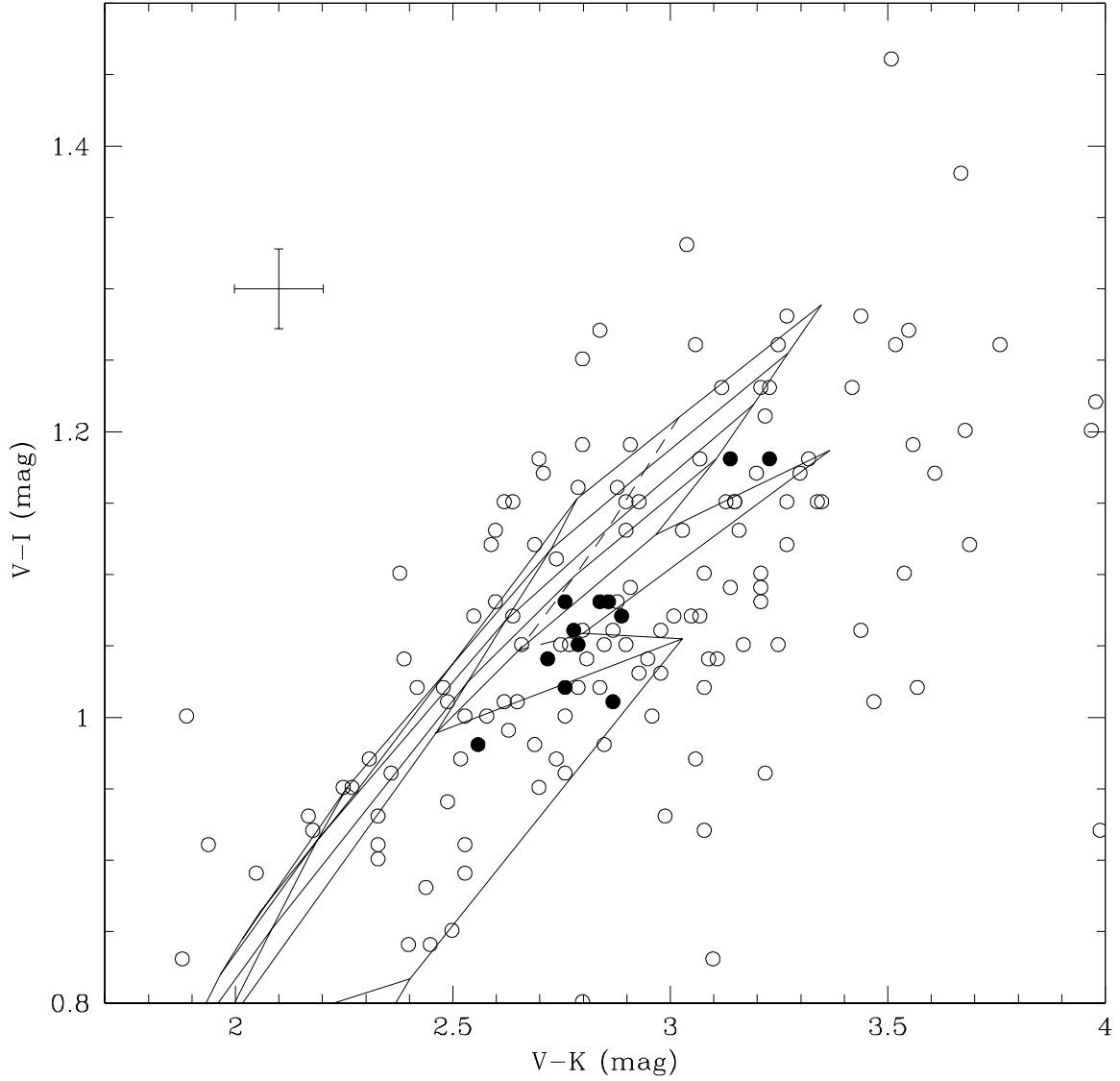}
\caption{$V-K$ vs.~$V-I$ color-color diagram for NGC 4365 GCs (circles). Filled circles are those GCs with spectroscopic data, either from this work or 
Paper I. Note our successful sampling of the region with candidate intermediate-age clusters. The median photometric error is shown. The overplotted model 
grid is from Maraston (2004). From bottom to top, the isochrone ages are 2, 3, 5, 8, 11, and 14 Gyr. From left to right, the isosiders (constant metallicity 
lines) are [$Z$/H] = $-1.33, -0.33, 0$  (dotted), and 0.33.}
\end{figure}
\clearpage  

\begin{figure}
\plotone{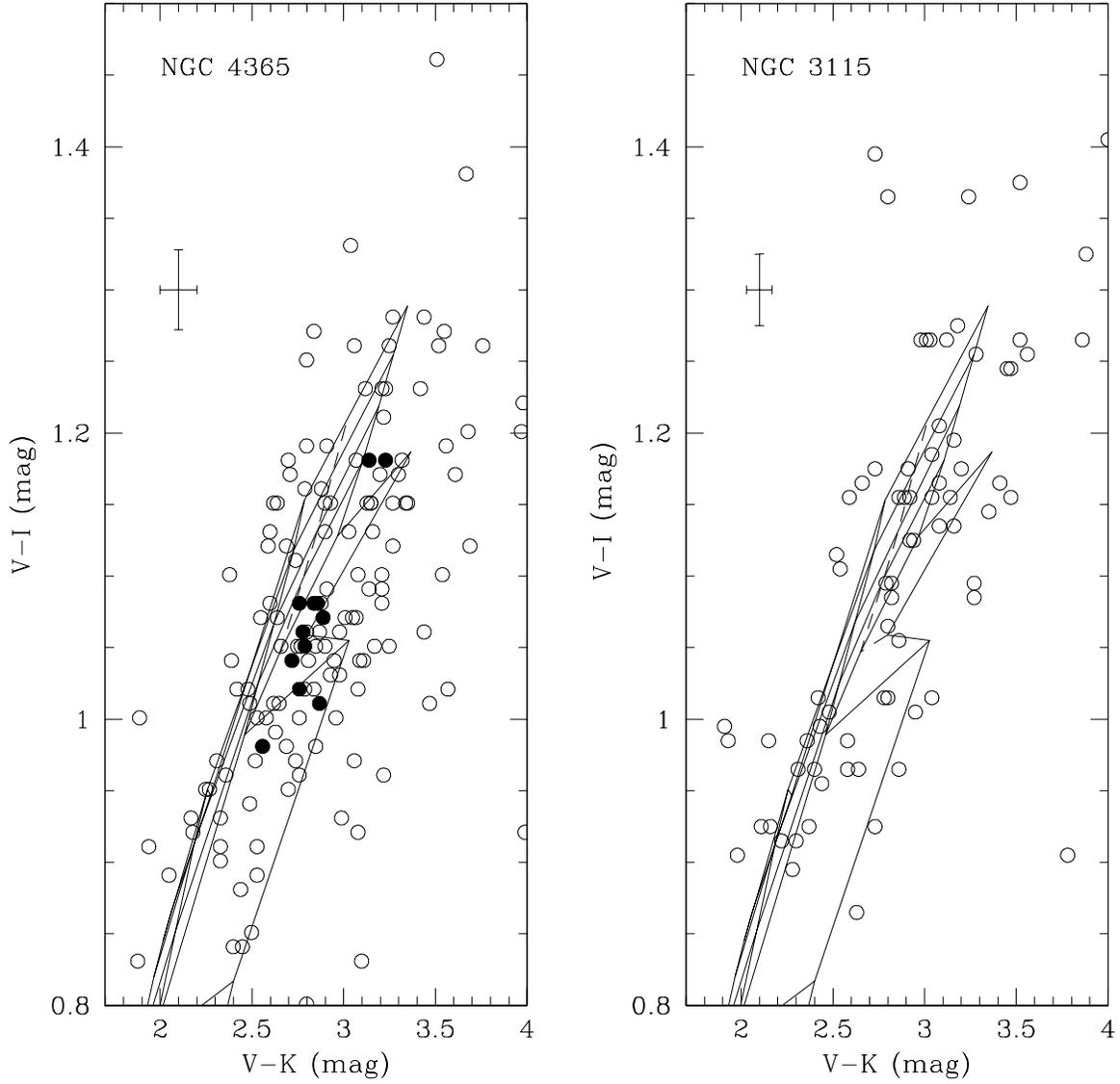}
\caption{$V-K$ vs.~$V-I$ color-color diagram for GCs in NGC 4365 and NGC 3115. The left panel is identical to the previous figure. The right panel clearly
shows the normal two GC subpopulations, which are not apparent in the NGC 4365 panel. Median photometric errors are shown. The model grids are from Maraston (2004), and grid identities are as in the previous figure.}
\end{figure}
\clearpage  

\begin{figure}
\plotone{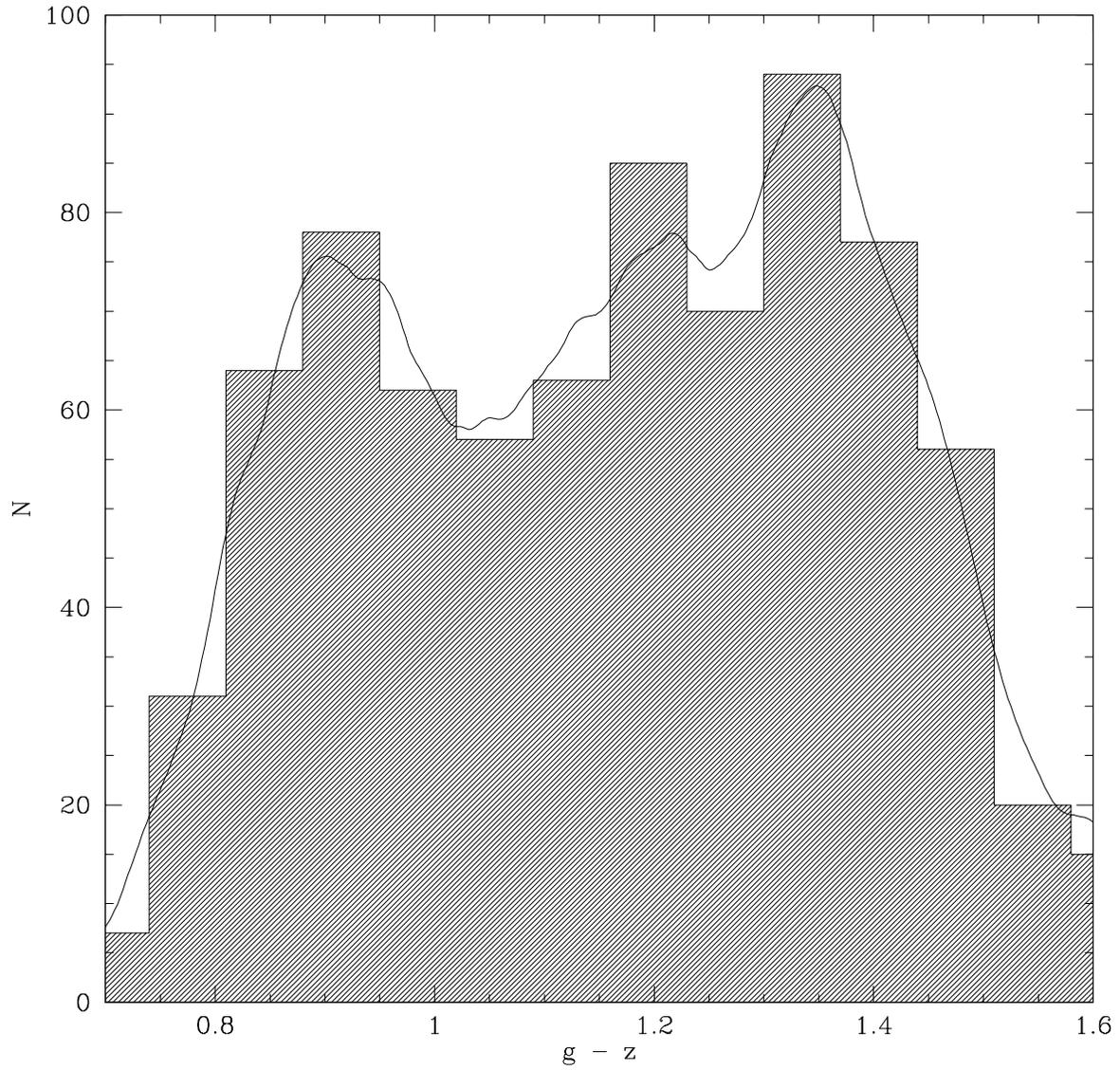}
\caption{$g-z$ color histogram for NGC 4365 clusters from HST/ACS. The solid line is a density estimate using an Epanechnikov kernel. The histogram appears trimodal, with 
peaks at $\sim 0.90, 1.22,$ and 1.34.}
\end{figure}
\clearpage  

\begin{figure}
\plotone{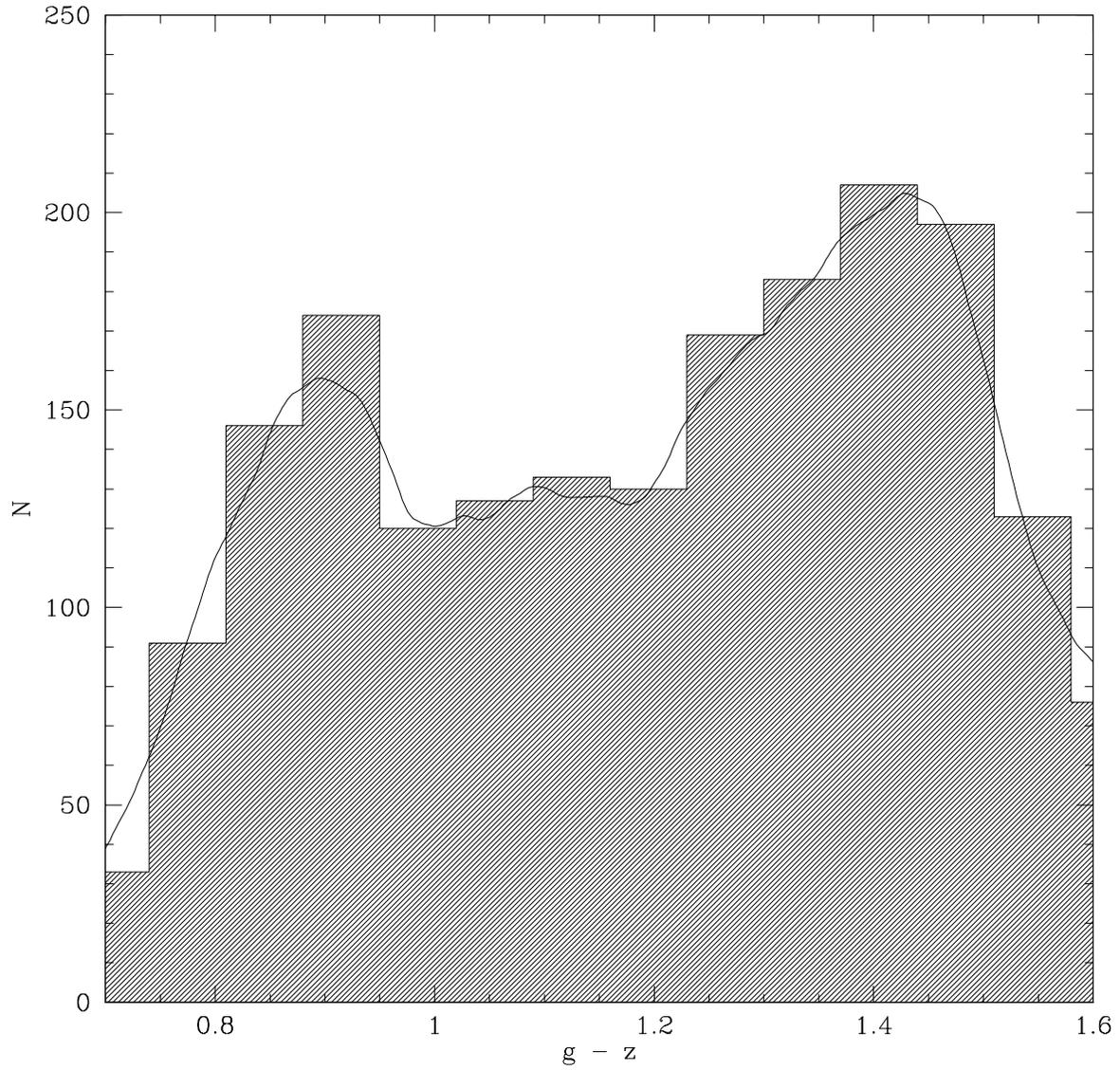}
\caption{$g-z$ color histogram for M87 clusters from HST/ACS. The solid line is a density estimate using an Epanechnikov kernel. The normal two subpopulations are apparent,  
at $\sim 0.89$ and 1.42.}
\end{figure}
\clearpage

\end{document}